\documentclass[conference,twocolumn,10pt]{IEEEtran} 
\usepackage[utf8]{inputenc}
\usepackage{epsfig,cite}
\usepackage{graphicx}
\graphicspath{{graphics/}}
\usepackage{color}
\usepackage{multirow} 
\usepackage{makecell} 
\usepackage{amssymb,amsmath}
\usepackage{mathrsfs} 

\IEEEoverridecommandlockouts

\DeclareMathOperator{\erf}{erf}

\begin{document}
\bibliographystyle{IEEEtran}

\title{Timing Control of Single Neuron Spikes with Optogenetic Stimulation}
\author{
	\IEEEauthorblockN{Adam Noel$^1$, Shayan Monabbati$^2$, Dimitrios Makrakis$^3$, Andrew W. Eckford$^4$}
	\IEEEauthorblockA{$^{1}$School of Engineering, University of Warwick, Coventry, UK, Email: adam.noel@warwick.ac.uk\\
		$^{2,4}$Department of EECS,
		York University,
		Toronto, Ontario, Canada, Email: $^2$shayan77@my.yorku.ca, $^4$aeckford@yorku.ca\\
	$^{3}$School of EECS, University of Ottawa, Ottawa, Ontario, Canada, Email: dimitris@eecs.uottawa.ca}
}

\newcommand{\EXP}[1]{\exp\left(#1\right)}
\newcommand{\ERF}[1]{\erf\left(#1\right)}

\newcommand{\metre}{\textnormal{m}}
\newcommand{\second}{\textnormal{s}}
\newcommand{\mol}{\textnormal{mol}}

\makeatletter
\newcommand{\vast}{\bBigg@{3}}
\newcommand{\Vast}{\bBigg@{4}}
\makeatother

\newcommand{\light}{\ell}
\newcommand{\rateTarget}{\lambda_\textrm{T}}
\newcommand{\fireThresh}{v_\textrm{fire}}
\newcommand{\tMin}{t_\textrm{min}}
\newcommand{\nMin}{{n_\textrm{min}}}
\newcommand{\tSep}{t_\Delta}
\newcommand{\nSep}{n_\Delta}
\newcommand{\distortion}{d}

\newcommand{\Imax}{I_\mathrm{max}}
\newcommand{\dt}{\Delta t}

\maketitle

\begin{abstract}
This paper predicts the ability to externally control the firing times of a cortical neuron whose behavior follows the Izhikevich neuron model. The Izhikevich neuron model provides an efficient and biologically plausible method to track a cortical neuron's membrane potential and its firing times. The external control is a simple optogenetic model represented by a constant current source that can be turned on or off. This paper considers a firing frequency that is sufficiently low for the membrane potential to return to its resting potential after it fires. The time required for the neuron to charge and for the neuron to recover to the resting potential are fitted to functions of the Izhikevich neuron model parameters. Results show that linear functions of the model parameters can be used to predict the charging times with some accuracy and are sufficient to estimate the highest firing frequency achievable without interspike interference.
\end{abstract}

\maketitle

\section{Introduction}

Over the past decade, developments in {\em optogenetics} have given researchers direct control over particular neurons \cite{Fenno2011,Deisseroth2011}.
Using this technique, neurons are modified with a gene that encodes a light-sensitive protein (an {\em opsin}), causing the neurons to express opsins on their surface. Certain opsins, such as channelrhodopsin \cite{Nagel2002}, open an ion channel in response to light. When the channels are open, an ion current flows through the neuron's membrane, changing its electrical potential and causing it to fire. Thus, if an optogenetically-modified neuron is stimulated with a strong light source, such as a laser, the neuron will fire in response. 

Dramatic advances in the study of the brain, as well as revolutionary new therapies for neurological disorders, are expected to follow from precise optogenetic control over neural circuits \cite{Deisseroth2012}. 
So far, research has often focused on the control of large groups of neurons in experimental settings \cite{Grosenick2015}; e.g., studies of seizures in the mouse brain \cite{Armstrong2013} or of spinal cord injury in rats \cite{Wenger2014}. However, targeted control of individual neural circuits are of considerable interest \cite{Rajasethupathy2016}. 

In this direction, an interesting problem is to precisely control the firing time of an individual neuron. 
Consider a neuron illuminated by a light source, where $i(t)$ is the time-varying light intensity. Let $\mathbf{t} = [t_1,t_2,\ldots,t_n]$ represent a vector of times at which the neuron fires. Then the neuron may be viewed as a functional $\mathsf{n}(\cdot)$, taking $i(t)$ as input and returning $\mathbf{t}$. The control problem is to invert $\mathsf{n}(\cdot)$: that is, given a desired vector $\mathbf{t}$, find $i(t)$ as a solution for $\mathbf{t} = \mathsf{n}(i(t))$.

The solution to this problem strongly depends on the neuron model $\mathsf{n}(\cdot)$, for which various models exist in the literature.
In our previous work \cite{Noel2017f,Noel}, we addressed the problem of finding $i(t)$ where $\mathsf{n}(\cdot)$ represents a discrete-time integrate-and-fire model \cite{Abbott1999}. The integrate-and-fire model considers neurons as capacitors, where the current is integrated over time to find the neuron's potential; once the potential exceeds a threshold, the neuron fires. Integrate-and-fire is simple and leads to tractable results, but hides much of the complexity of real neurons, motivating the search for alternatives. 

The main contribution of this paper is to understand how to control the optogenetic stimulation of neurons that follow the Izhikevich neuron model \cite{Izhikevich2003} (which we hereafter simply refer to as the \emph{Izhikevich model}). 
This model is relatively simple to describe and simulate, but is biologically plausible
because the range of neuron firing patterns that can be observed is consistent with all known types of cortical neurons, as demonstrated in \cite{Izhikevich2004} by tuning the model parameters. This is unlike other simple models, such as the integrate-and-fire model and its variants. The Izhikevich model has numerous possible modes of operation; for simplicity, we primarily focus on the location of a single firing time under \emph{Regular Spiking}. For various initial conditions, we perform a sensitivity analysis of the model and use curve fitting to estimate the illumination period required for the neuron to fire and recover. Our results show that our method leads to accurate control of the neuron, and (consistent with our results using the integrate-and-fire model) allows arbitrary spike sequences to be generated when there is sufficient time between consecutive spikes.

The rest of this paper is organized as follows. Section~\ref{sec_models} describes the optogenetic and membrane potential models. We couple the two models in Section~\ref{section_sim}. We fit the times for both charging and recovery, and demonstrate whether we can fire at target frequencies, in Section~\ref{section_fit}. We conclude in Section~\ref{sec_end}. 

\section{Physical Models}
\label{sec_models}

In this section, we briefly describe the two physical models that we integrate to describe the neuron stimulation and membrane potential. These are the optogenetic model for the external stimulation and the Izhikevich model for the membrane potential dynamics.

\subsection{Optogenetic System Model}

An optogenetically-modified neuron expresses light-sensitive proteins on its surface. A well-studied example is channelrhodopsin (ChR), which changes its shape in response to absorbing a photon; this shape change leads to the opening of an ion channel. In the channelrhodopsin-2 (ChR2) variant, the process of opening the channel has been modelled as a three-state continuous-time Markov chain \cite{Nagel2003}: ChR2 is initially in the {\em closed} state, with a closed ion channel. Absorbing a photon sends ChR2 into the {\em open} state, with an open ion channel. ChR2 relaxes into the {\em degraded} state, in which the ion channel is closed but the ChR2 cannot absorb a photon; and finally ChR2 returns to the {\em closed} state. 

While the ion channel is open, the ion current passing through the channel is dependent on a number of environmental factors, including pH and ion concentration \cite{Nagel2003}. Moreover, the dwell time in each state is a random variable, as is the number of receptors on the surface of the neuron. However, experimental results \cite{Nagel2002,Nagel2003} suggest that a neuron will experience a \emph{stable steady-state} current (on the order of $\mu$A or nA) in response to a constant illumination intensity $i(t)$. Thus, given a maximum current $\Imax$, we will assume there exists a known, deterministic mapping from $i(t)$ to current $I \in [0,\Imax]$. For the remainder of the paper, we will consider current $I$ rather than illumination $i(t)$.

\subsection{Izhikevich Neuron Model}

The Izhikevich model uses a two-dimensional system of ordinary differential equations having the membrane potential $v$ and the membrane recovery variable $u$ as variables. $u$, which accounts for the activation of potassium ionic current and the inactivation of sodium ionic currents, provides negative feedback to $v$. The system of equations was obtained via fitting to the spike initiation dynamics of cortical neurons and is as follows \cite[Eqs.~(1)--(3)]{Izhikevich2003}:
\begin{align}
\label{eqn_izhikevich_v}
    \frac{\mathrm{d}v}{\mathrm{d}t} = &\, 0.04v^2 + 5v + 140 - u + I, \\
\label{eqn_izhikevich_u}
    \frac{\mathrm{d}u}{\mathrm{d}t} = &\, a(bv-u), \\
    \textrm{if } v \geq &\, 30\,\textrm{mV, then}
    \left\{
	\begin{array}{l}
	v \leftarrow c \\
\label{eqn_izhikevich_reset}
	u \leftarrow u+d,
	\end{array}
	\right. 
\end{align}
where (\ref{eqn_izhikevich_v}) and (\ref{eqn_izhikevich_u}) update the rates of change of $v$ and $u$, respectively, and (\ref{eqn_izhikevich_reset}) resets $u$ and $v$ after a spike occurs. Time and potential are measured in $\mathrm{ms}$ and $\mathrm{mV}$, respectively. $I$ is the synaptic or input current. The parameters $a$, $b$, $c$, and $d$ are dimensionless parameters that can be tuned for different types of neurons; see Table~\ref{table_izhikevich_parameters}. $a$ sets the time scale of the decay of recovery variable $u$ after a spike occurs. $b$ describes the sensitivity of $u$ to subthreshold fluctuations of $v$. $c$ is the reset potential for $a$ after a spike occurs, and $d$ determines the reset of $u$ after a spike occurs.

Results in \cite{Izhikevich2003,Izhikevich2004} demonstrate that the Izhikevich model can produce the behaviors of different types of cortical neurons by appropriately tuning the parameters $\{a,b,c,d\}$, even though the model itself is not biophysically meaningful. Each type of neuron is associated with a characteristic firing pattern, where each firing pattern is a sequence of spikes. Our default parameter values in this work are consistent with \emph{Regular Spiking} (RS) neurons, which are the most typical neurons in the cortex. Their activity cannot be accurately represented with an integrate-and-fire model because they respond to a sustained stimulus by firing spikes with an interspike interval that is initially short and then increases. Also, when we calculate fitting functions in Section~\ref{section_fit}, we consider parameter ranges that include the other types of neurons listed in Table~\ref{table_izhikevich_parameters}.

\begin{table}[!t]
	\centering
	\caption{Selection of Nominal Parameter Values for the Izhikevich Neuron Model (from \cite{Izhikevich2003})}
	
	{\renewcommand{\arraystretch}{1.4}
		\begin{tabular}{l||c|c|c|c}
			\hline
			Neuron Type (Acronym) & $a$ & $b$ & $c$ & $d$ \\ \hline \hline
			Regular Spiking (RS) & 0.02 & 0.2 & -65 & 8 \\ \hline
			Fast Spiking (FS) & 0.1 & 0.2 & -65 & 2 \\ \hline
			Low-Threshold Spiking (LTS) & 0.02 & 0.25 & -65 & 2 \\ \hline
			Chattering (CH) & 0.02 & 0.2 & -50 & 2 \\ \hline
			Intrinsically Bursting (IB) & 0.02 & 0.2 & -55 & 4 \\ \hline
		\end{tabular}
	}
	\label{table_izhikevich_parameters}
\end{table}

\section{Simulating Neuron Spikes}
\label{section_sim}

In this section, we present the simulation of spikes in the Izhikevich model when it is stimulated by the simple optogenetic model. First, we describe the coupling of the simulation models and discuss the selection of suitable simulation parameters. We demonstrate the stimulation of a sequence of spikes and motivate our interest in studying individual spikes. Next, we study the impact of the model's initial conditions and derive the steady-state potentials of the Izhikevich model.

\subsection{Coupling the Izhikevich Model with Optogenetics}

We take a direct approach to couple the two physical models. We assume that there are no other current sources to the membrane and enable the optogenetic model to set the input current $I$ in (\ref{eqn_izhikevich_v}). The simple optogenetic model assumes that we have a binary current, which we can turn on and off as needed, thus we immediately have $I \in \{0,\Imax\}$. Thus, to simulate the complete system, we only need to initialize $\{u,v,I\}$ and use (\ref{eqn_izhikevich_v})--(\ref{eqn_izhikevich_reset}) in a loop to update $u$ and $v$, where we update $I$ or fire the neuron when required.

\begin{figure}[!t]
    \centering
    \includegraphics[width=\linewidth]{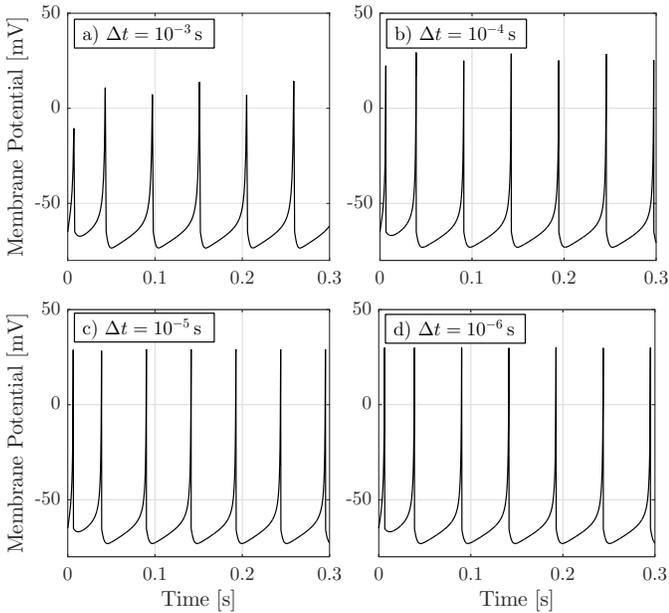}
    \caption{A sequence of neuron spikes for different values of time step $\dt$. The membrane is stimulated with a constant current $I=\Imax=10$. The model parameters are $\{a,b,c,d\} = \{0.02,0.2,-65,8\}$ (\emph{Regular Spiking} in Table~\ref{table_izhikevich_parameters}).}
    \label{fig_time_step}
\end{figure}

We must choose a time step $\dt$ to set the resolution with which we evaluate (\ref{eqn_izhikevich_v})--(\ref{eqn_izhikevich_reset}). In Fig.~\ref{fig_time_step}, we test different values of $\dt$ by setting the input current to a constant $I=\Imax=10$ (dimensionless). The default value of $\dt$ in \cite{Izhikevich2003,Izhikevich2004} is $\dt=10^{-3}\,\second$, but we see in Fig.~\ref{fig_time_step}a) that this results in an insufficient level of granularity. The apparent ``randomness'' in the membrane potential is not due to noise but are artefacts that can be mitigated by decreasing $\dt$. The timing of the spikes is indistinguishable for $\dt=10^{-5}\,\second$ and $\dt=10^{-6}\,\second$, and all their spikes peak at the same voltage (30\,mV). For this reason, we use $\dt=10^{-5}\,\second$ in the remainder of this work. We also always use $\Imax=10$.

From Fig.~\ref{fig_time_step}, we also observe that the interspike intervals are not constant, even when $\dt$ is sufficiently small. This behavior is expected for \emph{Regular Spiking} neurons and other types of neurons as well. However, our objective is to fit expressions to describe a neuron's behavior and control when it fires. In order to ignore the effects of interspike interference, we focus here on the generation of individual spikes.

\subsection{Initial Conditions and the Steady State}

To maintain accuracy in our numerical analysis, we need to impose consistent conditions on the membrane. To generate a single spike, we will turn the current ``on'' until the neuron fires and then leave the current ``off''. In the absence of an input current, the membrane potential of a neuron should converge to a \emph{resting potential} (unless it is bistable or inhibition induced; see \cite{Izhikevich2004}). By setting the LHS of (\ref{eqn_izhikevich_v}) and (\ref{eqn_izhikevich_u}) to $0$, and the input $I$ to $0$, we can show that the two possible resting potentials are
\begin{equation}
\label{eqn_rest_potential}
    v_\mathrm{rest} = 12.5b - 62.5 \pm 12.5 \sqrt{b^2 -10b + 2.6}.
\end{equation}

The smaller solution of (\ref{eqn_rest_potential}), $v_\mathrm{rest}^-$, is stable. The larger solution $v_\mathrm{rest}^+$ is unstable and is in fact the firing threshold. If the membrane potential is higher than $v_\mathrm{rest}^+$, then $v$ will increase even if $I=0$ and the neuron will fire (though firing could be avoided with sufficiently large \emph{negative} current). If the membrane potential is lower than $v_\mathrm{rest}^+$ and no input is applied, then the potential will converge to $v_\mathrm{rest}^-$. Throughout this work, we assume that the voltage has converged once it remains within $\epsilon=0.5\%$ of $v_\mathrm{rest}^-$.

\begin{figure}[!t]
    \centering
    \includegraphics[width=\linewidth]{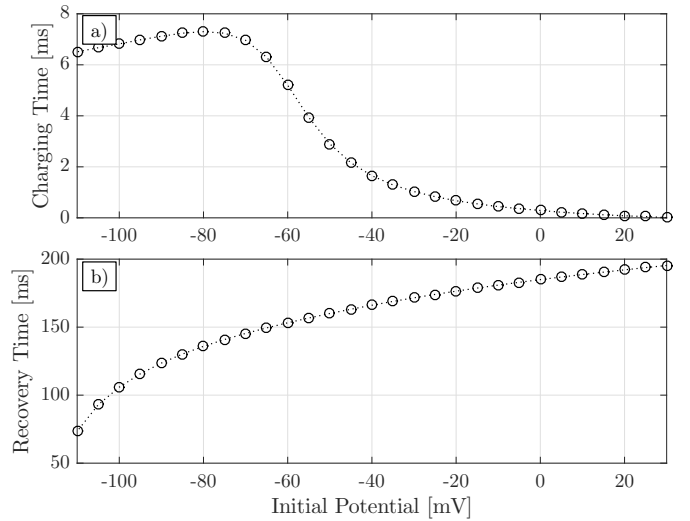}
    \caption{a) Charging time and b) recovery time as a function of the initial membrane potential. The input current $I$ remains on until the neuron fires. The model parameters are $\{a,b,c,d\} = \{0.02,0.2,-65,8\}$ (\emph{Regular Spiking} in Table~\ref{table_izhikevich_parameters}).}
    \label{fig_behavior_vs_initial}
\end{figure}

We refer to the time needed for the neuron to fire as the \emph{charging time} and the time to reach the stable resting potential as the \emph{recovery time}. We show in Fig.~\ref{fig_behavior_vs_initial}, where $v_\mathrm{rest}^-=-70\,\mathrm{mV}$, that \emph{both} of these times are sensitive to the initial membrane potential. To make this work relevant to the generation of multiple spikes, we impose that the initial membrane potential is also the resting potential $v_\mathrm{rest}^-$, and that the recovery variable $u$ is initially $bv$ (i.e., (\ref{eqn_izhikevich_u}) is 0).

\section{Fitting Models for Timing Behavior}
\label{section_fit}

In this section, we assess whether we can predict the timing behavior, i.e., the charging and recovery times of the Izhikevich neuron model, based on knowledge of the model parameters. Specifically, we seek numerically-derived equations for a neuron's behavior as a function of $\{a,b,c,d\}$. We are not predisposed towards any particular class of equations, but we seek results that are sufficiently accurate to use as a guide to control firing times and know how long to wait between firing times (for the membrane to return to the resting potential before charging again). Our assumptions limit the usefulness of very high precision; the optogenetic model is simplified, the model parameters cannot be directly measured, and we do not consider noise sources. Nevertheless, we seek to gain intuition about controlling a neuron, and in particular we will be able to estimate the maximum firing frequency that can be achieved without interspike interference.

The remainder of this section is as follows. First, we measure the charging time and the recovery time as functions of individual model parameters, where the remaining model parameters are fixed. This helps us decide which parameters to focus on in a joint model. The charging time only depends on $a$ and $b$ (as expected from (\ref{eqn_izhikevich_v})--(\ref{eqn_izhikevich_reset})), and the recovery time is most sensitive to $a$ and $d$. Next, we measure the charging time as a function of both $a$ and $b$ and the recovery time as a function of both $a$ and $d$. All fitting functions are found via nonlinear least squares in MATLAB. Finally, we demonstrate that we can estimate the maximum firing frequency that can be achieved without interference between spikes. We show that deviations from target firing times can occur when we try to fire the neuron more frequently. Unless otherwise noted, the default parameter values are the nominal values for a \emph{Regular Spiking} neuron, i.e., $\{a,b,c,d\} = \{0.02,0.2,-65,8\}$, as shown in Table~\ref{table_izhikevich_parameters}.

We measure the accuracy of the fitting functions with three methods. $R^2$ measures the proportion of the variance in the behavior that is predictable from the model parameters, where $R^2 \in [0,1]$. The root mean square error (RMSE) measures the standard deviation of the behavior from that predicted by the fitting functions. The maximum error (Max Error) is simply the absolute value of the largest deviation from the fitting function over the parameter range or ranges considered.

\subsection{Fitting to Individual Model Parameters}

To fit the behavior to the individual model parameters, we vary one parameter while holding the remaining parameters constant. The chosen range of each parameter is in consideration of the types of neurons listed in Table~\ref{table_izhikevich_parameters}. We consider polynomial functions up to degree 2 (i.e., linear and quadratic), and exponential functions with either 1 or 2 terms. These are the simplest fitting functions in MATLAB's Curve Fitting Toolbox. The range of each varied parameter, a selection of fitted equations for their behavior (chosen for space), and the accuracy of each fit are summarized in Table~\ref{table_fitting_one_parameter}.

\begin{table*}[!t]
	\centering
	\caption{Fitting neuron behavior to a single parameter. Default values are $\{a,b,c,d\} = \{0.02,0.2,-65,8\}$.}
	{\renewcommand{\arraystretch}{1.4}
		\begin{tabular}{l|c||c|c|c|c|c}
			\hline
			Parameter & Range & Behavior & Function & $R^2$ & RMSE [ms] & Max Error [ms] \\ \hline \hline
			\multirow{3}{*}{$a$} & \multirow{3}{*}{[0.01,0.14]} & Charging & $2a+6.92$ & $1.0$ & $1.11\times10^{-15}$ & $1.78\times10^{-15}$ \\ \cline{3-7}
			& & \multirow{2}{*}{Recovery} &
			$360.3e^{-35.22a}$ & $0.8996$ & $21.62$ & $32.79$ \\ \cline{4-7}
			& & & $564e^{-105a}+98.23e^{-10.74a}$ & $0.9992$ & $1.896$ & $3.059$ \\ \hline
			\multirow{4}{*}{$b$} & \multirow{4}{*}{[0.1,0.25]} &  \multirow{2}{*}{Charging} & $1034b^2 - 463.4b + 57.37$ & $0.9540$ & $1.1128$ & $2.9522$ \\ \cline{4-7}
			& & & $6255e^{-64.5b} + 29.62e^{-7.224b}$ & $0.9998$ & $0.0691$ & $0.1653$ \\ \cline{3-7}
			& & \multirow{2}{*}{Recovery} & $-1255b^2+454b+103.7$ & $0.6818$ & $1.6848$ & $4.377$ \\ \cline{4-7}
			& & & $-3.631\times10^{-6}e^{61.13b}+131e^{0.5454b}$ & $0.9990$ & $0.0951$ & $0.2061$ \\ \hline
			\multirow{2}{*}{$c$} & \multirow{2}{*}{[-80,-50]} & \multirow{2}{*}{Recovery} & $0.06786c+150$ & $0.9104$ & $0.2128$ & $0.3807$ \\ \cline{4-7}
			& & & $147.4e^{2.21\times10^{-4}c} + 2729e^{0.1539c}$ & $0.9999$ & $0.0076$ & $0.0110$ \\ \hline
			\multirow{2}{*}{$d$} & \multirow{2}{*}{[2,8]} & \multirow{2}{*}{Recovery} & $7.708d+86.79$ & $0.9742$ & $2.5090$ & $4.1525$ \\ \cline{4-7}
			& & & $121.7e^{0.02502d}-62.69e^{-0.3712d}$ & $1.0$ & $0.0386$ & $0.0585$ \\ \hline
		\end{tabular}
	}
	\label{table_fitting_one_parameter}
\end{table*}

\begin{figure}[!t]
    \centering
    \includegraphics[width=\linewidth]{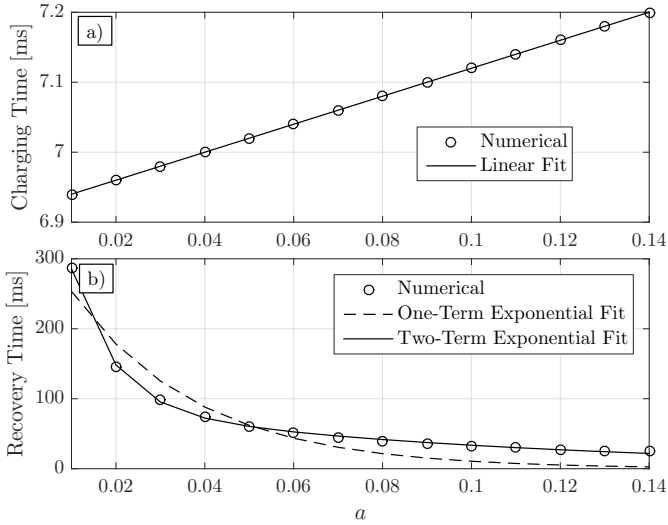}
    \caption{a) Charging time and b) recovery time as a function of $a$. The input current remains on until the neuron fires.}
    \label{fig_a_behavior}
\end{figure}

We show the neuron behavior as functions of $a$ in Fig.~\ref{fig_a_behavior}. As $a$ increases, the charging time increases whereas the recovery time decreases. The charging time only varies by less than $0.3\,\mathrm{ms}$ and is accurately represented by a linear function. The two-term exponential fit is better suited for the recovery time and agrees very well ($R^2=0.9992$).

\begin{figure}[!t]
    \centering
    \includegraphics[width=\linewidth]{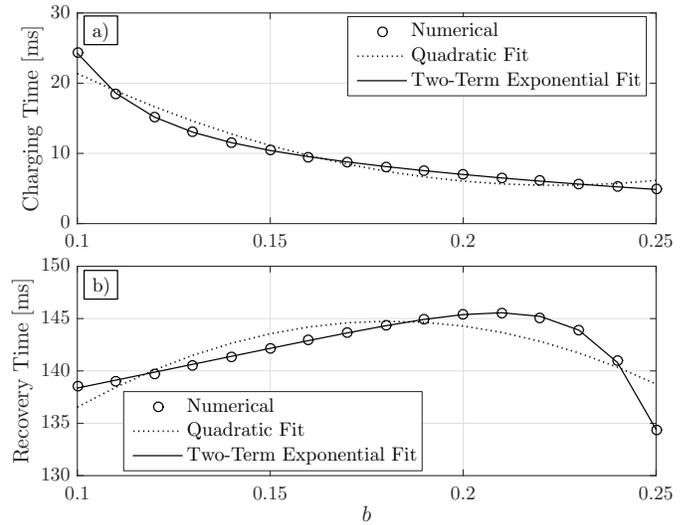}
    \caption{a) Charging time and b) recovery time as a function of $b$. The input current remains on until the neuron fires.}
    \label{fig_b_behavior}
\end{figure}

We show the neuron behavior as functions of $b$ in Fig.~\ref{fig_b_behavior}. As $b$ increases, the charging time decreases, whereas the recovery time increases and then later decreases. The charging time can be matched with a quadratic function ($R^2=0.9540$), although the two-term exponential fit is much stronger. Only the two-term exponential fit is suitable for the recovery time. Nevertheless, we note that the relative range of the recovery time as a function of $b$ is somewhat small (with less than $8\%$ deviation from the largest value).

\begin{figure}[!t]
    \centering
    \includegraphics[width=\linewidth]{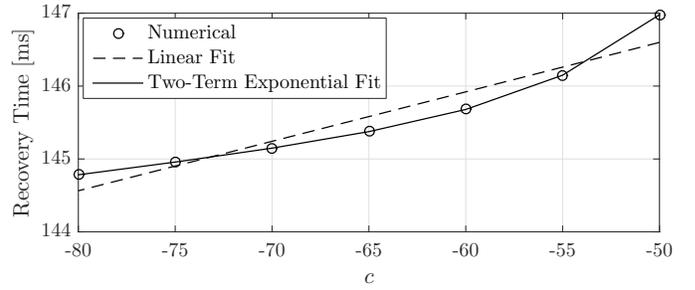}
    \caption{Recovery time as a function of $c$. The input current remains on until the neuron fires.}
    \label{fig_c_recovery}
\end{figure}

From (\ref{eqn_izhikevich_v})--(\ref{eqn_izhikevich_reset}), we know that $c$ and $d$ have no impact on the charging time \emph{when the neuron is only fired once}. We show the neuron recovery time as a function of $c$ in Fig.~\ref{fig_c_recovery}, where we see that the recovery time increases with $c$. The two-term exponential fit is very accurate ($R^2=0.9999$), but even the linear fit has a maximum error of only $0.3807\,\mathrm{ms}$. The total range in the recovery time is about $2.25\,\mathrm{ms}$, or about $1.5\%$ deviation from the largest value.

\begin{figure}[!t] 
    \centering
    \includegraphics[width=\linewidth]{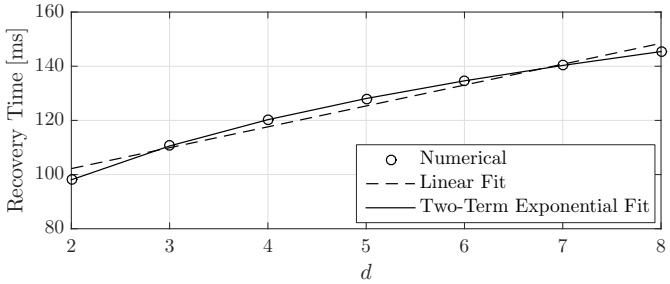}
    \caption{Recovery time as a function of $d$. The input current remains on until the neuron fires.}
    \label{fig_d_recovery}
\end{figure}

We show the neuron recovery time as a function of $d$ in Fig.~\ref{fig_d_recovery}. As $d$ increases, the recovery time also increases. Once again, the two-term exponential fit is the most accurate ($R^2=1.0$), but the linear fit also agrees well ($R^2=0.9742$).

\subsection{Fitting to Multiple Model Parameters}

\begin{figure}[!t]
    \centering
    \includegraphics[width=\linewidth]{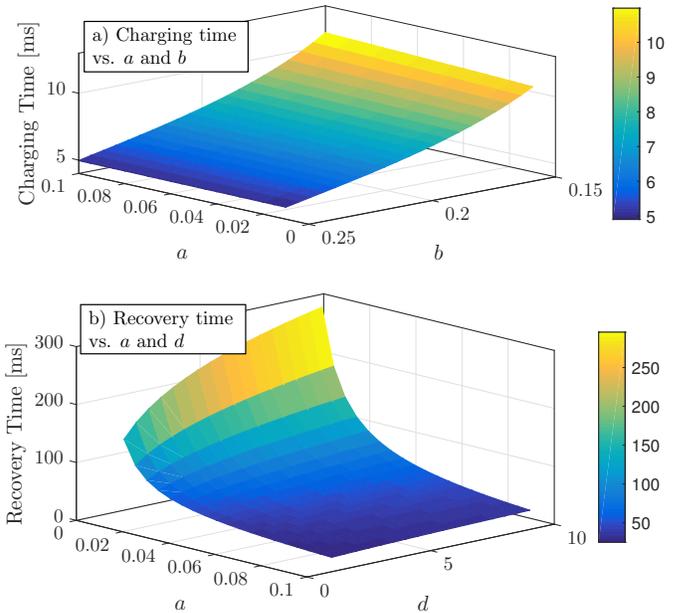}
    \caption{Neuron spiking behavior as a function of multiple model parameters. a) shows charging time as a function of $a$ and $b$, and b) shows recovery time as a function of $a$ and $d$ where $b=0.2$ and $c=-65$.}
    \label{fig_abd_behavior}
\end{figure}

The previous results enable us to predict the parameters that are most suitable for multiple-parameter models of the neuron behavior. Clearly, only $a$ and $b$ are relevant for the charging time. The charging time is much more sensitive to $b$ than to $a$, but we still include both because the magnitude of the charging time is relatively much smaller than the recovery time. All four parameters are relevant for the recovery time, but in the interest of simplicity we choose to fit according to $a$ and $d$ while keeping $b=0.2$ and $c=-65$, since the recovery time is less sensitive to $b$ and $c$. In Fig.~\ref{fig_abd_behavior}, we plot the charging time as a function of both $a$ and $b$ and the recovery time as a function of both $a$ and $d$. Two-parameter fitting functions for charging time and recovery time are listed for these parameter ranges in the first four rows of Table~\ref{table_fitting_multi_parameters}. We consider both linear and quadratic surfaces.

\begin{table*}[!t]
	\centering
	\caption{Fitting neuron behavior for multiple parameters over different ranges of $a$, $b$, and $d$, where $c=-65$}
	
	{\renewcommand{\arraystretch}{1.4}
		\begin{tabular}{l||c|c|c|c|c|c|c}
			\hline
			Behavior & $a$ & $b$ & $d$ & Function & $R^2$ & \thead{RMSE \\ $[\mathrm{ms}]$} & \thead{Max Error \\ $[\mathrm{ms}]$} \\ \hline \hline
			\multirow{2}{*}{Charging} & \multirow{2}{*}{[0.01,0.1]} & \multirow{2}{*}{[0.15,0.25]} & \multirow{2}{*}{--} & $18.09 + 2.2621a - 54.69b$ & $0.9754$ & $0.2630$ & $0.8276$ \\ \cline{5-8}
			& & & & $29.46 + 13.45a - 174.1b - 1.973a^2 -53.07ab + 305.8b^2$ & $0.9983$ & $0.0692$ & $0.2265$ \\ \hline
			\multirow{2}{*}{Recovery} & \multirow{2}{*}{[0.01,0.1]} & \multirow{2}{*}{0.2} & \multirow{2}{*}{[1,9]} & $136.6-1572a+4.07d$ & $0.6568$ & $31.94$ & $138.0$ \\ \cline{5-8}
			& & & & $181.1-5082a + 14.55d + 37460a^2 - 122ad -0.3766d^2$ & $0.8914$ & $17.97$ & $72.07$ \\ \hline
			\hline
			\multirow{2}{*}{\makecell{Charging \\ (FS)}} & \multirow{2}{*}{[0.08,0.12]} & \multirow{2}{*}{[0.15,0.22]} & \multirow{2}{*}{--} & $20.19 + 3.096a - 66.52b$ & $0.9794$ & $0.2083$ & $0.5137$ \\ \cline{5-8}
			& & & & $35.51 +16.15a - 241.6b - 3.261a^2 -67.05ab + 491.4b^2$ & $0.9993$ & $0.0393$ & $0.0949$ \\ \hline
			\multirow{2}{*}{\makecell{Recovery \\ (FS)}} & \multirow{2}{*}{[0.08,0.12]} & \multirow{2}{*}{0.2} & \multirow{2}{*}{[1,3]} & $43.41 - 196.3a + 1.519d$ & $0.9832$ & $0.3538$ & $0.9573$ \\ \cline{5-8}
			& & & & $56.39 -523.7a + 4.643d + 1894a^2 -25.71ad -0.1381d^2$ & $0.9997$ & $0.0446$ & $0.0987$ \\ \hline
		\end{tabular}
	}
	\label{table_fitting_multi_parameters}
\end{table*}

We see in Table~\ref{table_fitting_multi_parameters} that the fitting functions for the charging time have good accuracy, even in the linear case ($R^2=0.9754$ with all errors less than $1\,\mathrm{ms}$). However, the accuracy in the recovery time prediction is relatively poor, with all fits having $R^2<0.9$ and a high RMSE. The reason for this is that we are fitting to the recovery time over a large range for both $a$ and $d$, corresponding to the neuron types listed in Table~\ref{table_izhikevich_parameters}. To improve on this, we consider constraining the parameter ranges to correspond to a specific type of neuron, i.e., \emph{Fast Spiking} (FS) neurons. The fitting functions for this type are also listed in Table~\ref{table_fitting_multi_parameters}. The expressions for the recovery time are much more accurate than they were for the larger parameter ranges, e.g., the linear fit for the recovery time of the FS neuron has $R^2=0.9832$, with all errors less than $1\,\mathrm{ms}$.

\subsection{Multiple Neuron Spikes}

\begin{figure}[!t]
    \centering
    \includegraphics[width=\linewidth]{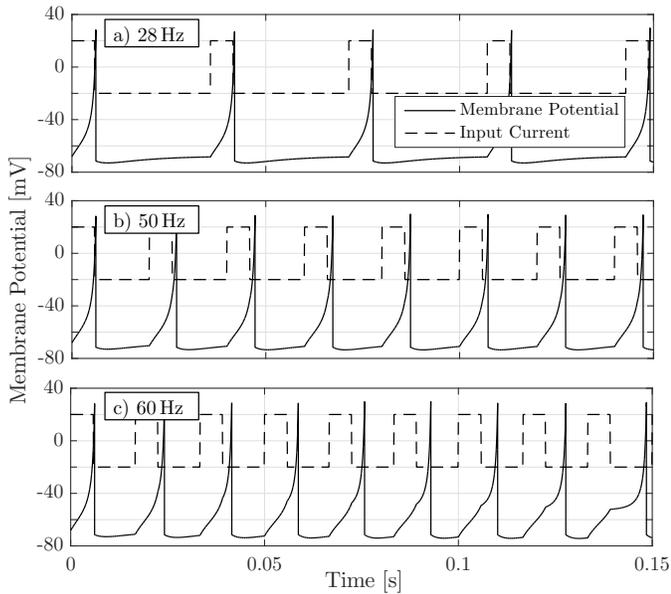}
    \caption{Membrane potential versus time for a modified FS neuron ($\{a,b,c,d\} = \{0.09,0.22,-71.5,2.2\}$) that is stimulated to fire at a specified frequency. From the linear fit in Table~\ref{table_fitting_multi_parameters}, the minimum period is $34.92\,\mathrm{ms}$, i.e., the maximum interference-free frequency is $28.6\,\mathrm{Hz}$. The current is drawn on an arbitrary scale to show when it turns on and off.}
    \label{fig_freq_match}
\end{figure}

Finally, we apply our numerical analysis to estimate the maximum firing frequency of a neuron. We choose model parameters that are $10\%$ from the nominal values of a FS neuron, where $\{a,b,c,d\} = \{0.09,0.22,-71.5,2.2\}$. Both $a$, $b$, and $d$ are within the ranges of values used for the FS fitting functions in Table~\ref{table_fitting_multi_parameters}, although we assumed for the recovery time that $\{b,c\}=\{0.2,-65\}$. From the fitting functions, the charging and recovery times for the FS neuron are estimated to be $5.837\,\mathrm{ms}$ and $29.08\,\mathrm{ms}$, respectively. This gives a total period of $34.92\,\mathrm{ms}$. Let us consider the accuracy of this fit by trying to generate spikes with a frequency of $28\,\mathrm{Hz}$, such that the target period between spikes is $35.71\,\mathrm{ms}$. We do this by repeatedly turning the current on for $5.837\,\mathrm{ms}$ and then off for $(35.71-5.837)=29.873\,\mathrm{ms}$, \emph{whether or not the neuron has actually fired}. We can then observe whether the neuron fired after $5.837\,\mathrm{ms}$, and then every $35.71\,\mathrm{ms}$ after that. We observe the results in Fig.~\ref{fig_freq_match}a), where we see that the neuron keeps firing just as the current is turned off.

If we increase the target frequency to $50\,\mathrm{Hz}$, as in Fig.~\ref{fig_freq_match}b), the neuron does not recover to the resting potential before the current comes back on. Thus, the neuron cannot keep up with the target rate. We see this even more clearly with large delays in Fig.~\ref{fig_freq_match}c) by increasing the target frequency to $60\,\mathrm{Hz}$. We conclude that our numerical analysis enables us to estimate the maximum (interference-free) firing frequency of a neuron.

\section{Conclusion}
\label{sec_end}

In this paper, we have considered the use of optogenetic stimulation to control the timing of individual neuron spikes. We used the Izhikevich model for the neuron membrane potential dynamics and fitted the neuron behavior to functions of the model's parameters. We have demonstrated that linear functions were sufficient to predict the highest firing frequency that can be achieved in a \emph{Fast Spiking} neuron without interspike interference. Future work will measure the deviations that occur when a neuron does not fire when specified (e.g., as we considered for integrate-and-fire neurons in \cite{Noel2017f, Noel}).

\section*{Acknowledgment}

This work was supported in part by the Natural Sciences and Engineering Research Council of Canada (NSERC).
    
\bibliography{NoelEckfordDistortion}

\end{document}